\pdfoutput=1
\documentclass[twocolumn,pra,aps,showpacs,secnumarabic]{revtex4-1}
\usepackage{graphicx}
\usepackage{color}

\usepackage{epsfig}
\usepackage{epstopdf}
\usepackage[T1]{fontenc}
\usepackage{bm}

\usepackage{amsmath}

\newcommand{\beq}{\begin{eqnarray}}
\newcommand{\eeq}{\end{eqnarray}}

\begin{document}
	\setcounter{page}{1}
	
	\title{ Trapped Bose-Bose mixtures at finite temperature: a quantum Monte Carlo approach}
	\author{ K. D\v{z}elalija$^1$,  V. Cikojevi\'c$^{1,2}$, J. Boronat$^2$, L. Vranje\v{s} 
		Marki\'{c}$^1$}
	\affiliation{$^1$Faculty of Science, University of Split, Ru\dj era 
		Bo\v{s}kovi\'{c}a 33, HR-21000 Split, Croatia}
	\affiliation{$^2$Departament de F\'{\i}sica, Campus Nord B4-B5, Universitat Polit\`ecnica de Catalunya, E-08034 Barcelona, Spain}
	
	\begin{abstract}
           We study thermal properties of a trapped Bose-Bose mixture in a dilute regime using quantum Monte Carlo methods.  
        Our main aim is to investigate the dependence of the superfluid density and the condensate fraction on temperature, for the mixed and separated phases. To this end, we use the diffusion Monte Carlo method, in the zero-temperature limit, and the path-integral Monte Carlo method for finite temperatures. The results obtained are compared with solutions of the coupled Gross-Pitaevskii equations for the mixture at zero temperature. We notice the existence of an anisotropic superfluid density in some phase-separated mixtures. Our results also show that the temperature evolution of the superfluid density and condensate fraction is slightly different, showing noteworthy situations where the superfluid fraction is smaller than the condensate fraction.
	\end{abstract}


	\maketitle

	\section{Introduction}
Mixtures of quantum fluids offer the possibility to study effects arising from the interaction of their components, mainly on  fundamental phenomena such as superfluidity and Bose-Einstein condensation.  The naturally occurring fermion-boson $^3$He-$^4$He mixture has for many decades been the focus of such studies, since naturally occurring bosonic mixtures were not available.  Much more versatile have proven to be degenerate two-component bosonic systems of ultracold atoms, which can be achieved by trapping two hyperfine components of the same species~\cite{Myatt97,Hall98,Maddaloni00}, different isotopes~\cite{Papp08,Sugawa11,Stellmer13} or different elements~\cite{Modugno02,Thalmahhmer08,Lercher11,McCarron11,Pasquiou13,Wacker15,Wang16,Schulze18,Trautmann18,Burchianti18}.  The possibility to change both the interactions of their components, by Feshbach resonances, and the dimensionality of the system has provided a very rich landscape of physical phenomena.

Mixtures of repulsive ultracold Bose gases in harmonic traps can be created in different regimes, mixed or phase-separated, depending on the inter- and intra-particle interactions and the mass ratio of the 
components~\cite{ho,pu,ohberg,Kean16, Cikojevic18}.  The phase separation can occur in two ways: by separation in two blobs or by one species being pushed further from the center and forming a shell around the other. The extreme diluteness of the trapped gases enables a description of the interatomic interactions using only one parameter, the s-wave scattering length $a$. The mean-field predictions for the miscibility agree rather well with quantum Monte Carlo predictions, especially when the strength of the interactions is small~\cite{Cikojevic18}, but the density profiles can be quite different. The existence of a trap and the atom numbers can modify the miscibility criteria~\cite{Kean16}. However, the character of the system (miscible or phase separated) remains universal when one considers systems with the same Gross-Pitaevskii scaling parameter $Na/l_{{\rm ho}}$, with $l_{{\rm ho}}=\sqrt{\hbar/(m\omega)}$ the oscillator length~\cite{Cikojevic18}.

In recent years, Bose-Bose mixtures with attractive interspecies and repulsive intraspecies interactions have attracted considerable interest, due to the theoretical prediction~\cite{petrov} and posterior realization~\cite{bb_mixture_first_taruell,soliton_to_drop,bb_mixture_sec}
of  ultradilute liquid droplets.
The stability of these droplets is a result of a delicate balance between mean-field and its first perturbative correction, the Lee-Huang-Yang term.    Recent studies show that even corrections to this term are necessary to explain some experimental phenomena  and that the inclusion of finite effective ranges in the theoretical analysis allows for a better understanding of these droplets~\cite{cikojevic,staudinger,symmetric, tononi1,tononi2,salsnich_nonuniversal,cikojevic20}.

In the limit of zero temperature, single component ultradilute alkali gases are nearly 100\% condensed and the superfluidity accompanies Bose-Einstein condensation (BEC). In the same limit, liquid $^4$He  is fully superfluid, but only around 7\% of the atoms are condensed~\cite{Glyde18} due to strong interparticle interactions.  In the field of cold atoms, it is  experimentally possible to reach quantum degeneracy for both components of either trapped gases or quantum droplets. This enables investigation of the effects of one superfluid on the other, as well as the relationship of  condensate and superfluid fractions in the limit of zero and finite temperature.

In this paper, we study both superfluidity and Bose-Einstein condensation of trapped repulsive Bose-Bose mixtures in an exact way using Quantum Monte Carlo methods, at both zero and finite temperature. Although it would be very interesting to study attractive mixtures as well, the large number of particles required for realistic mixed droplets ($>10^4$) makes their present study from first principles not feasible.  The rest of the paper is organized as follows. In Section 2, we introduce the chosen models and the theoretical methods used for the study.  Section 3 reports, for three representative regimes, results for the superfluid and condensate fractions as a function of temperature. Finally, in Section 4 we discuss the main findings of our study.

\section{Model and method}

We investigated a balanced Bose-Bose mixture with number of particles $N_1 = N_2=N$ and masses $m_1 = m_2 = m$, trapped in a three-dimensional harmonic confinement with a common frequency $\omega$. The Hamiltonian of the system is
\begin{eqnarray}
H  & = & -  \frac{\hbar^2}{2m} \sum_{i=1}^{2N} \nabla_i^2 + \frac{1}{2} m \omega^2 \sum_{i=1}^{2N} r_i^2 \\ \nonumber 
  &  & + \frac{1}{2}   
\sum_{\alpha, \beta=1}^{2} \sum_{i_\alpha, j_\beta=1}^{N, 
N}V^{(\alpha, \beta)}(r_{i_\alpha j_\beta}) \ ,
\label{hamiltonian}
\end{eqnarray}
where $V^{(\alpha, \beta)}(r_{i_\alpha j_\beta})$ is the interatomic 
potential between species $\alpha$ and $\beta$. We employed short-range interatomic potentials in the form
\begin{equation}
V^{(\alpha, \beta)}(r) = u^{(\alpha, \beta)} \, \left( \frac{r_0^{(\alpha, \beta)}}{r}     \right)^{10} \ .
\label{potential}
\end{equation}
The parameters of the interaction potential ($u$, $r_0$) were chosen to reproduce the desired scattering length. Given these two parameters, there are numerous ways to construct an interaction potential for a given s-wave scattering length. Particularly, we chose $r_{0}^{(\alpha, \beta)} = 1.5 a_{\alpha \beta}$, where $a_{\alpha \beta}$ is the scattering length in the $\alpha \beta$ channel, and numerically solved the two-body Schr\"odinger equation to find a corresponding strength of the potential $u^{(\alpha, \beta)}$~\cite{scattering}. In previous work it was verified that, in the conditions of the present simulations, the specific shape of the interaction potential is not relevant because we are in a universal regime in terms of the gas parameter~\cite{Cikojevic18}.

At zero temperature, we  used the second-order diffusion Monte Carlo method (DMC), which solves stochastically the Schr\"odinger equation written in imaginary time, as described in Ref.~\cite{Boronat:94a}. The trial wave function was modeled as
\begin{eqnarray}
\Psi(\mathbf{R}) & = &
    \prod_{1=i<j}^{N_1} f^{(1,1)}(r_{ij})
\prod_{1=i<j}^{N_2}f^{(2,2)}(r_{ij}) \prod_{i,j=1}^{N_1, 
N_2}f^{(1,2)}(r_{ij}) \nonumber \\
            & & \times \prod_{i=1}^{N_1} h^{(1)}(r_i)
		\prod_{j=1}^{N_2} h^{(2)}(r_j) \ ,
\label{trialw}
\end{eqnarray}
where $\mathbf{R} = \left\{ \mathbf{r}_1, \ldots , \mathbf{r}_N \right\}$, 
$h^{(i)}$ is the one-body term, due to the external harmonic potential, and 
$f^{(\alpha, \beta)}$ is a two-body Jastrow factor which accounts for 
two-particle correlations. We chose $f^{(\alpha, \beta)}$ to be a solution of 
the two-body problem for a given potential, and $h^{(\alpha)} = \exp( - r^2 / (2 
c_\alpha^2))$ ($\alpha=1,2$),  $c_\alpha$ being a variational parameter. 
Specific values of $c_\alpha$ depend on the interaction potential and, in all 
the cases, we found them to be very close to $l_{\rm ho}$, with $l_{\rm ho}$  the 
corresponding harmonic length, because of the diluteness of the system. In all 
the presented DMC results we used $c_\alpha = l_{\rm ho}$. All possible 
systematic errors, in particular time-step and population-size biases were 
thoroughly investigated and reduced below the statistical noise. 
The estimation of observables which do not commute with the Hamiltonian was 
carried out using forward-walking pure estimators~\cite{pures} 
We  used about 
300 walkers and a time step of around $ 10^{-3} m a_{11}^2 / \hbar^2$ for shell 
and mixed regimes, and 10000 walkers for a two-blobs regime.

We also compared our DMC results with the solutions of coupled Gross-Pitaevskii equations for quantum mixtures~\cite{ho},
\begin{eqnarray}
            i \hbar \frac{\partial \phi_{1} (\mathbf{r},t)}{\partial t}
           & = &
            \left(
            -\frac{\hbar^2}{2 m}\nabla^2
            +
            \dfrac{1}{2}m \omega^2 r^2
            +
            g_{11} |\phi_{1}(\mathbf{r},t)|^2  \right.  \nonumber \\
         & &    +       g_{12} |\phi_{2}(\mathbf{r},t)|^2
            \bigg) 
            \phi_{1} (\mathbf{r},t) \ ,
\label{gp1}
\end{eqnarray}
\begin{eqnarray}
	i \hbar \frac{\partial \phi_{2} (\mathbf{r},t)}{\partial t}
	& = &
	\left(
	-\frac{\hbar^2}{2 m} \nabla^2
	+
	\dfrac{1}{2}m \omega^2 r^2
	+
	g_{22} |\phi_{2}(\mathbf{r},t)|^2 \right. \nonumber \\
 & &	+
	g_{12} |\phi_{1}(\mathbf{r},t)|^2
	\bigg) 
	\phi_{2} (\mathbf{r},t) \ ,
\label{gp2}
\end{eqnarray}
where 
\begin{equation}
    g_{ij} = \dfrac{2\pi \hbar^2 a_{ij}}{\mu} = 2\pi \hbar^2a_{ij} \left(m_i^{-1} + m_j^{-1}\right)
\end{equation}
are the interaction strengths. 
These equations were solved simultaneously by imaginary time propagation of $\psi = (\phi_1, \phi_2)$
\begin{equation}
    \psi(t + \Delta t) = e^{-H\Delta t} \psi(t) \ ,
\end{equation}
where a Trotter decomposition of the $\mathcal{O}(\Delta t^3)$ for the time evolution operator was used~\cite{chin_krotchek},
\begin{equation}
    e^{-H\Delta t} = e^{- \Delta tV(\mathbf{r})/2} e^{-\hbar \Delta t \nabla^2  / (2m) } e^{- \Delta tV(\mathbf{r}')/2} + \mathcal{O}(\Delta t^3) \ ,
\end{equation}
$V=(V_1, V_2)$ being the potential acting on each of the components,
\begin{eqnarray}
V_1  & = &  \dfrac{1}{2}m \omega^2 r^2 + g_{11} |\phi_1|^2 + g_{12}|\phi_2|^2, \\
V_2  & = &  \dfrac{1}{2}m \omega^2 r^2 + g_{22} |\phi_2|^2 + g_{12}|\phi_1|^2 \ .
\end{eqnarray}
Along the time evolution, the kinetic energy propagator $e^{-\hbar \Delta t \nabla^2  / (2m) }$ was evaluated in Fourier space by means of the fast Fourier transform.

At finite temperature, we carried out simulations using the path integral Monte Carlo (PIMC) method including the worm algorithm to better sample the permutation space~\cite{Boninsegni:06,Boninsegni:06a}. This is an exact method if the number of terms of every path $M$ (\textit{beads}) is large enough to observe the convergence guaranteed by the Trotter theorem. In our simulations, we got convergence working in a range of imaginary-time steps between $0.00036$ and $0.001$ $(0.5\hbar\omega)^{-1}$.

\begin{figure}[tb]
	\centering
	\includegraphics[width=8cm]{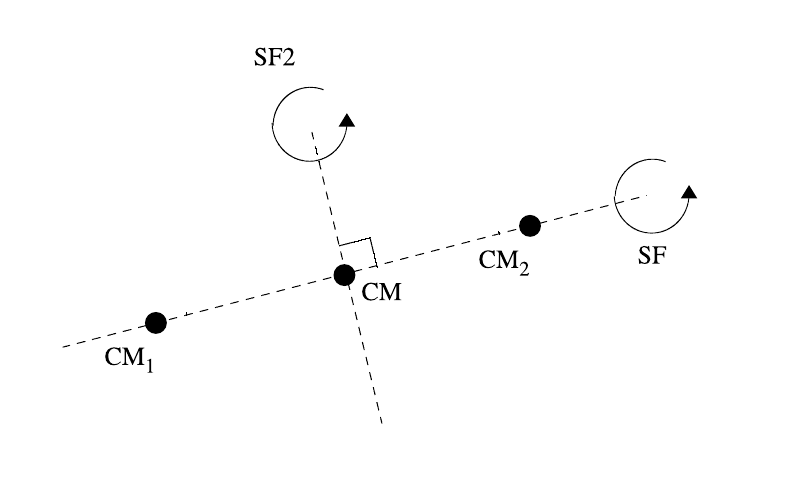}
	\caption{Sketch of the two axes around which the superfluid fraction is 
calculated, for the two separated components $B_1$ and $B_2$. CM$_1$ (CM$_2$) 
stands for the center-of-mass of component 1 (2) and CM stands for the total 
center-of-mass.}
	\label{fig:axis}
\end{figure}

The superfluid fraction of each component, as a function of temperature, was estimated in PIMC using the area estimator~\cite{Ceperley89}, 
\begin{equation}
\frac{\rho_S}{\rho_0}=1-\frac{I}{I_C}=\frac{2m\langle A_z^2\rangle}{\beta \lambda I_C} \  , 
\label{area}
\end{equation}
where $\lambda=\frac{\hbar^2}{2m}$ and ${\bf A}=\frac{1}{2}\sum_{i=1}^N\sum_{j=1}^M {\bf r}_{i,j}\times {\bf r}_{i,j+1},$. In Eq. (\ref{area}), $I$ is the moment of inertia of the system, with respect to a given rotation axis, and $I_c$ is its classical value. The $z$ component of ${\bf A}$
can be understood as a projection area of all polymers onto a plane perpendicular to the rotation axis.
In order to calculate the superfluid fraction of each component (\ref{area}) in the trapped mixture, the moment of inertia $I$ was calculated with respect to axis passing through the center of mass. Two different axes were used, as shown in Fig. \ref{fig:axis}, where it is assumed that the  centers of mass of the two species  are separated.

The Bose-Einstein condensation in the trapped system was studied by calculating and diagonalizing the one-body density matrix of each component, in terms of the natural orbitals~\cite{Lowdin,Lewart,DuBois}. The condensate fraction was obtained as the eigenvalue corresponding to the orbital with the highest occupation.

\section{Results}

Bose-Bose mixtures have a very rich phase diagram driven by the interplay 
between the strength of the three different pair interactions.  We have chosen 
to study three representative points in the phase space, that were previously 
explored at zero temperature~\cite{Cikojevic18} and are thus expected to 
correspond to two blobs, mixed and shell phase. According to the mean field 
approach~\cite{Kean16}, an adimensional parameter 
$\Delta=g_{11}g_{22}/g_{12}^2-1$ classifies the regime of phase separation 
($\Delta<0$) and miscibility ($\Delta>0$) in a homogeneous system at $T=0$. At 
finite temperatures, a homogeneous mixture can phase separate even for $\Delta >0$ due to the effects 
of Bose-Einstein condensation \cite{ota2019magnetic}. In a harmonically trapped 
Bose mixture~\cite{Cikojevic18,Kean16}, the complete phase diagram at $T=0$ 
is additionally characterized by the trapping length and number of particles. 
The boundaries between the three observed regimes (two blobs, mixed and shell) 
are not well defined, and here we only studied representative points of the phase 
diagram. According to $T=0$ results~\cite{Cikojevic18,Kean16},
two blobs are expected to occur with 
interaction strengths ratios $g_{12}/g_{22}=3$ and $g_{11}/g_{12}=0.33$ 
($\Delta=-0.89)$, mixed phase at $g_{12}/g_{22}=0.33$ and $g_{11}/g_{12}=3$ 
($\Delta=8$), and shell phase at 
$g_{12}/g_{22}=3$ and $g_{11}/g_{12}=1.7$ 
($\Delta=-0.43$). Those coordinates correspond to 
the points A, F, and G in Ref. \onlinecite{Cikojevic18}, respectively.
Our system is composed of 
$N_1=N_2=50$ atoms, and in all three cases we took $m_1 = m_2 = m$.  In a 
previous study~\cite{Cikojevic18}, it was shown that this configuration is in 
the  universal regime, where the results do not depend on the details of the 
interaction potential but only on the scattering lengths. This was concluded by 
performing DMC calculations with two interatomic potentials which shared the 
same value of the s-wave scattering length: i) a 10-6 Lennard-Jones-like 
potential~\cite{pade} , and ii) a hard-core potential. In this work, instead of 
performing calculations with a hard-core potential, we chose a purely repulsive 
$\sim 1/r^{10}$ potential to model the interaction in all three channels
because it was more suitable for the employed  implementation of the PIMC 
method. The parameters of the potential were adjusted so that the  two-body 
bound state is not supported. Additionally, it was shown that the shape of the 
density profiles remained the same, with just the norm changing accordingly, 
when the calculations were performed for $N =200$ and $400$ atoms, provided that 
$N_1 a_{11} / l_{\rm ho}$, $N_1 a_{12} / l_{\rm ho}$, $N_2 a_{12} / l_{\rm ho}$ 
and $N_2 a_{22} / l_{\rm ho}$ are kept fixed. This scaling allows comparison of 
theoretical predictions with experimental results without simulating directly 
large-$N$ systems using time-consuming QMC methods. In fact, this universality 
emerges from the Gross-Pitaevskii equations (Eqs. 
\ref{gp1} and \ref{gp2}) when they are written in  
length, energy, and time scales given by $l_{\rm ho} = \sqrt{\hbar / (m 
\omega)}$, $\hbar^2/(m l_{\rm ho}^2)$ and $\tau =m l_{\rm ho}^2 / \hbar $, 
respectively
\begin{eqnarray}
i  \frac{\partial \tilde{\phi}_{1} (\tilde{\mathbf{r}},\tilde{t})}{\partial \tilde{t}}
& = &
\left(
-\frac{\tilde{\nabla}^2}{2}
+
\dfrac{1}{2}\tilde{r}^2
+
\dfrac{4 \pi N_1 a_{11}}{l_{\rm ho}} |\tilde{\phi}_{1}(\tilde{\mathbf{r}},\tilde{t})|^2  \right.  \nonumber \\
& &    +       \dfrac{4 \pi N_2 a_{12}}{l_{\rm ho}} |\tilde{\phi}_{2}(\tilde{\mathbf{r}},\tilde{t})|^2
\bigg) 
\tilde{\phi}_{1} (\tilde{\mathbf{r}},\tilde{t}) \ ,
\label{gp1_red}
\end{eqnarray}
\begin{eqnarray}
i  \frac{\partial \tilde{\phi}_{2} (\tilde{\mathbf{r}},\tilde{t})}{\partial \tilde{t}}
& = &
\left(
-\frac{\tilde{\nabla}^2}{2}
+
\dfrac{1}{2}\tilde{r}^2
+
\dfrac{4 \pi N_2 a_{22}}{l_{\rm ho}} |\tilde{\phi}_{2}(\tilde{\mathbf{r}},\tilde{t})|^2  \right.  \nonumber \\
& &    +       \dfrac{4 \pi N_1 a_{12}}{l_{\rm ho}} |\tilde{\phi}_{1}(\tilde{\mathbf{r}},\tilde{t})|^2
\bigg) 
\tilde{\phi}_{2} (\tilde{\mathbf{r}},\tilde{t}) \ ,
\label{gp2_red}
\end{eqnarray}
where $\tilde{\phi}_1$ and $\tilde{\phi}_2$ are normalized according to $\int 
d^3\tilde{r} |\tilde{\phi}_i|^2 = 1$ ($i=1,2$). Notice that the universality is 
recovered when fixing the ratios $g_{12} / g_{22}$, $g_{11} / g_{12}$ and $N_1 
a_{11}/l_{\rm ho}$ since in our study $m_1=m_2$ and $N_1=N_2$.

\begin{table}[]
	\centering
	\label{table:scattering_params}
	\caption{Parameters corresponding to the mixed, two-blobs, and 
shell phase points analyzed ($N_1 = N_2 = 50$). Particles are trapped with 
harmonic confinement of length $l_{\rm ho}=\sqrt{\hbar / (m \omega)}$ (note that 
we use $m_1=m_2=m$). 
We also report $\Delta=g_{11}g_{22}/g_{12}^2-1$ and $\delta g / g$, where $g = 
\sqrt{g_{11} g_{22}}$ and $\delta g = \sqrt{g_{11} g_{22}} - g_{12}$.}
	\begin{tabular}{c | c | c | c | c | c | c | c} \hline
		& $\frac{g_{11}}{g_{12}}$ & $\frac{g_{12}}{g_{22}}$ & $\Delta$  & $\delta g / g$ &  $\frac{N_1 a_{11}}{l_{\rm ho}}  $ &  $\frac{N_2 a_{22}}{l_{\rm ho}}$  & $ \frac{N_1 a_{12}}{l_{\rm ho}} $ \\[+0.3em] \hline
		Mixed regime  & 3                 & 0.33  & 8            & 0.67           & 5               & 5            & 1.67            \\
		Two blobs     & 0.33              & 3 & -0.89  & -2          & 5               & 5            & 15           \\
		Shell         & 1.7               & 3 & -0.43 & -0.33          & 5               & 0.98            & 2.94                                           
	\end{tabular}
\end{table}

In the following, we present our results for each of the chosen regimes.

\subsection{Mixed Regime}

\begin{figure}[h!]
	\centering
	\includegraphics[width=8cm]{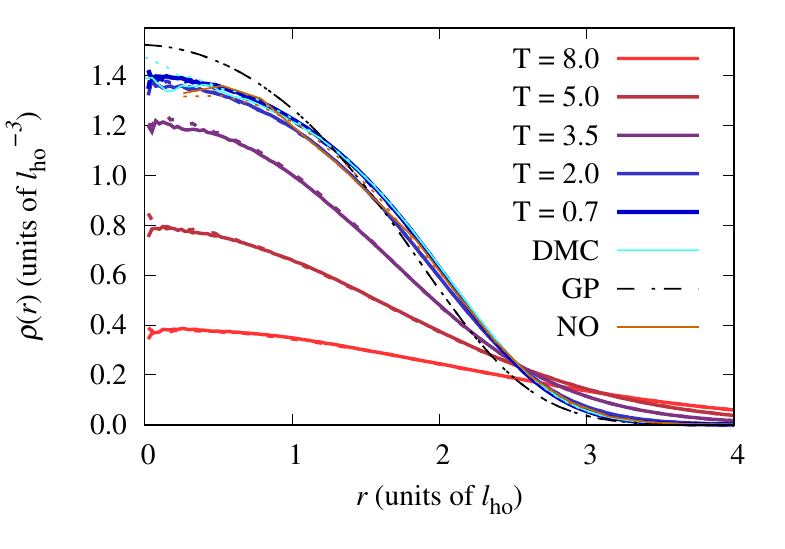}
		\caption{Density profiles for species 1 ($B_1$) (dashed lines) and
species 2 ($B_2$) (solid lines) centered at the minimum of the external 
potential in the mixed regime (parameters given in Table \ref{table:scattering_params}). The temperature is given in units of 
$\hbar\omega/2$. The results at finite temperatures are obtained by PIMC. NO 
represents the natural orbital for $T = 0.7$.}
		\label{mixgr}
\end{figure}

The density profiles of both species are presented in Fig.~\ref{mixgr}, both at zero temperature, from DMC and GP calculations and at finite temperatures, from PIMC. Temperatures are given in units $\hbar\omega/2$. The profiles of both species  coincide within the statistical errorbars, as expected. GP predicts a slightly narrower distribution than DMC. For the temperature $T=0.7$ PIMC and DMC results coincide within errorbars, while with the rise of temperature the profile becomes wider. The natural orbital with macroscopic occupation for $T=0.7$ coincides with the corresponding density profile.

\begin{figure}[t!]
	\centering
	\includegraphics[width=8cm]{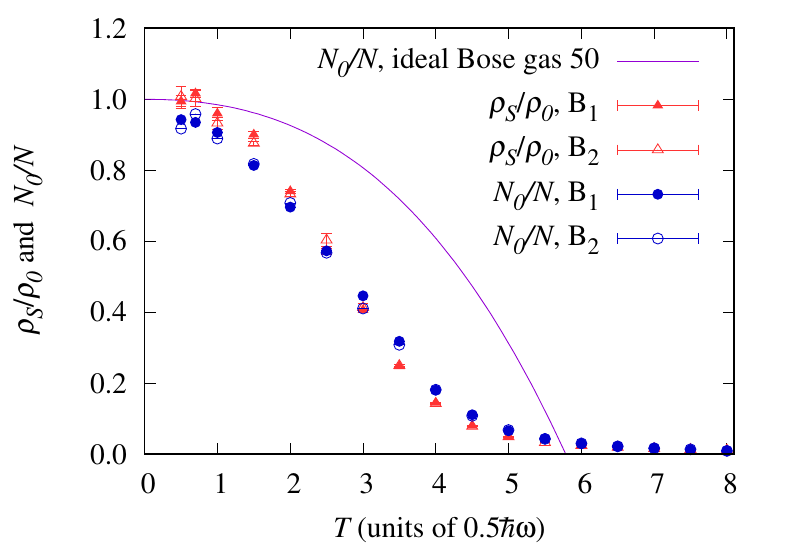}
	\caption{Superfluid and condensate fraction as a function of temperature for $B_1$ and $B_2$ in the mixed regime (parameters given in Table \ref{table:scattering_params}). The solid thin line stands for the condensate fraction for an ideal confined Bose gas. Temperature is given in units $\hbar\omega/2$.}
	\label{mixsf}
\end{figure}

Fig. \ref{mixsf}  reports the superfluid and condensate fractions as a function of temperature for both components $B_1$ and $B_2$. In the limit of zero temperature, the superfluid fraction is one, while the condensate fraction is slightly lower due to the interactions. With the increase of temperature both fractions decrease reaching zero simultaneously. The observed decay of the condensate fraction is significantly different from the one for an ideal Bose gas~\cite{dalfovo}, in the same conditions as the mixture, due to the depletion induced by dynamic correlations. Interestingly, around $T=3$ the condensate fraction becomes slightly larger than the superfluid fraction. We verified that this feature also happens for a single species in the same conditions, and that it disappears progressively when the strength of the interaction increases. 
The unusual regime in which the condensate fraction is larger than the superfluid fraction was previously predicted in conditions of very strong disorder~\cite{Astrakharchik02}.

\subsection{Separated regime: two blobs}
\begin{figure}[h!]
	\centering
	\includegraphics[width=8cm]{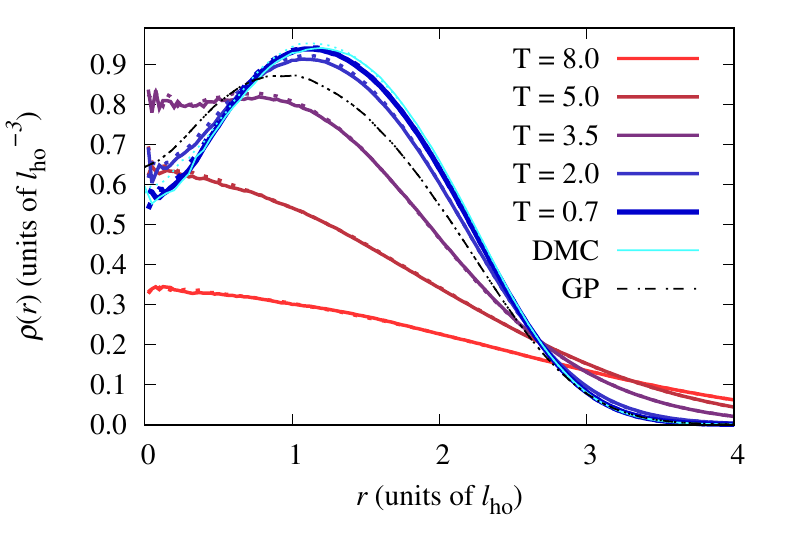}
	\caption{Density profiles in the two-blobs phase (parameters given in Table \ref{table:scattering_params}) for $B_1$ (dashed) and $B_2$ (solid) estimated from the minimum of the harmonic potential. }
	\label{blobsgr}
\end{figure}

\begin{figure}[h!]
	\centering
	\includegraphics[width=8cm]{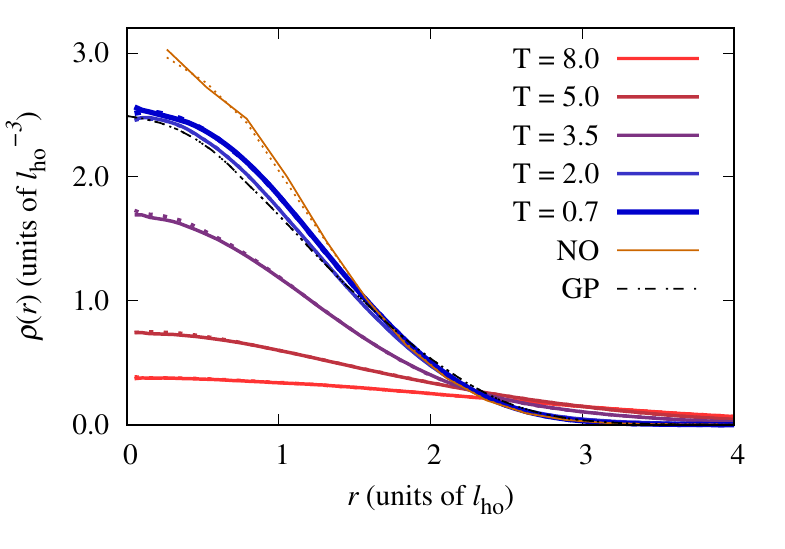}
	\caption{Density profiles in the two-blobs phase (parameters given in Table \ref{table:scattering_params}) for $B_1$ (dashed) and $B_2$ (solid) estimated from the center of mass of each component.}
	\label{blobsgr2}
\end{figure}

\begin{figure}[h!]
	\centering
	\includegraphics[width=8cm]{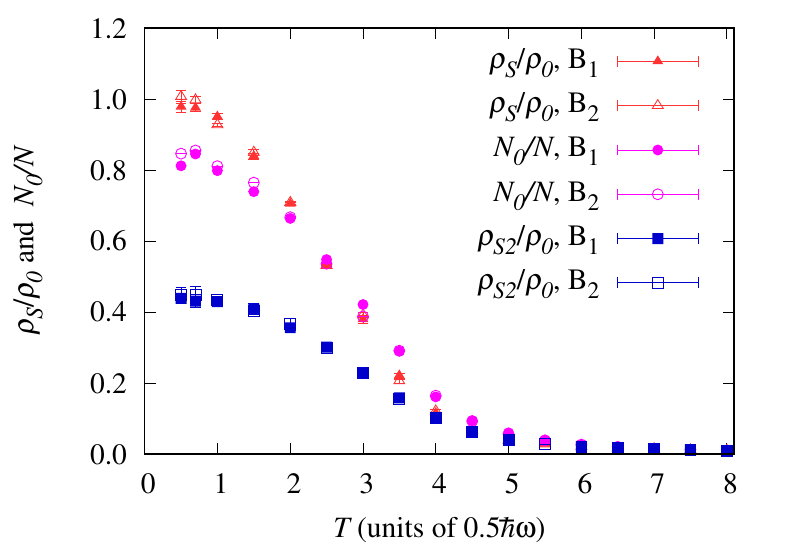}
		\caption{Superfluid and condensate fractions as a function of temperature for $B_1$ and $B_2$ in the two-blobs phase (parameters given in Table \ref{table:scattering_params}). }
		\label{blobssf}
\end{figure}

In this separated regime, the two species appear as forming a two blob 
structure~\cite{Cikojevic18}. The density profiles, estimated from the center of 
the harmonic well, are presented in Fig.~\ref{blobsgr}. Both the results at 
$T=0.5$ and 0.7 are within the errorbars equal to the zero-temperature DMC 
profiles and demonstrate separation in two blobs. With the increase of 
temperature, the blobs increasingly mix. At low temperatures, GP profiles do not 
fully match with QMC ones. As both QMC methods fully capture many-body 
correlations, the disagreement in the density profiles originates from  
quantum fluctuations, which are relevant since we obtain inner 
densities $\rho_1 a_{11}^3 \simeq 10^{-3}$. Since in the regime of two blobs, 
the LHY term is imaginary~\cite{petrov,Hu20p}, the QMC approach stands as a favorable 
method to investigate beyond mean-field effects in this phase. Additionally, 
in Fig.~\ref{blobsgr2} we report the density profiles and natural orbitals 
estimated from the center of mass of each component. Both density profiles 
remain nearly unchanged up to $T=2$, and then start to spread. The 
macroscopically occupied natural orbital follows the density profile except at 
short distances, $r<1.5$.
Due to the separation of the centers of masses of the two species, the axis up to which superfluid estimator is calculated using the area estimator is important. 
In Fig.~\ref{blobssf}, we report the superfluid and condensate fractions as a 
function of temperature for both components $B_1$ i $B_2$.  For the superfluid 
fraction two results are given: SF which represents rotation around the axis 
which passes through both components' center of mass and SF2 which passes 
through the total center of mass and is perpendicular to the previous axis, as shown 
in Fig.~\ref{fig:axis}. It is worth noticing that this separated phase shows 
an anisotropic superfluid response, feature already observed for instance in  
in Helium drops with impurities~\cite{sungjin2015, shinichi2007}.
In the limit of zero temperature, SF goes to one while SF2 is significantly lower, below 0.5, which can be explained by the drag produced by the other component. In this same limit, the condensate fraction is in between the two superfluid values, around 0.85, which is lower than in the case of the mixed regime. As the temperature increases, after a certain temperature, the condensate fraction becomes larger than both superfluid fractions, just like in the mixed regime.

\subsection{Separated regime: shell}
\begin{figure}[h!]
	\centering
	\includegraphics[width=8cm]{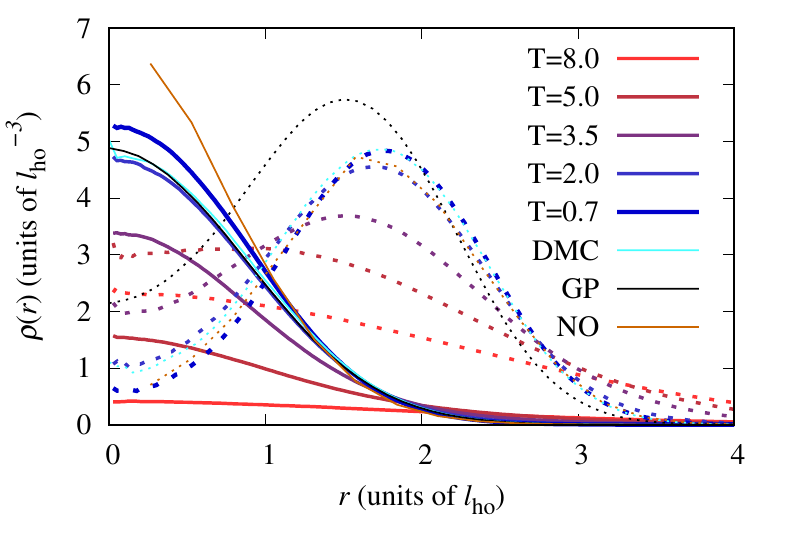}
		\caption{Density profiles in the shell phase (parameters given in Table \ref{table:scattering_params}) for $B_1$ (dashed) and $B_2$ (solid) estimated 
		from the minimum of the harmonic potential. Notice the different scale for the two components, $\rho_{B_1}/\rho_{B_2}=7$.}
		\label{shellgr}
\end{figure}

In this case, the mixture is also separated in two parts but now there is spherical symmetry with respect to the trap center; one species occupies the inner part and the other forms a spherical shell surrounding the inner one. 
The density profiles of the mixture in this regime are presented in 
Fig.~\ref{shellgr}. The agreement between GP and QMC solutions at zero 
temperature is only qualitative. As in the two-blob configuration, this is most 
likely to occur due to quantum fluctuations, since here the peak density 
is $\rho_1 a_{11}^3 \simeq 5 \times 10^{-3}$, beyond the limit of 
applicability of the mean-field theory. 
Species 1 ($B_1$) remains well in the outer shell up to temperatures $T\simeq 2$ 
and then it starts to partially mix with the inner species 2 ($B_2$). The 
dominant natural orbital for the  $B_1$ component is close to the density 
profile at low temperature, while the one for $B_2$ is more peaked at small 
distances.

\begin{figure}[h!]
	\centering
	\includegraphics[width=8cm]{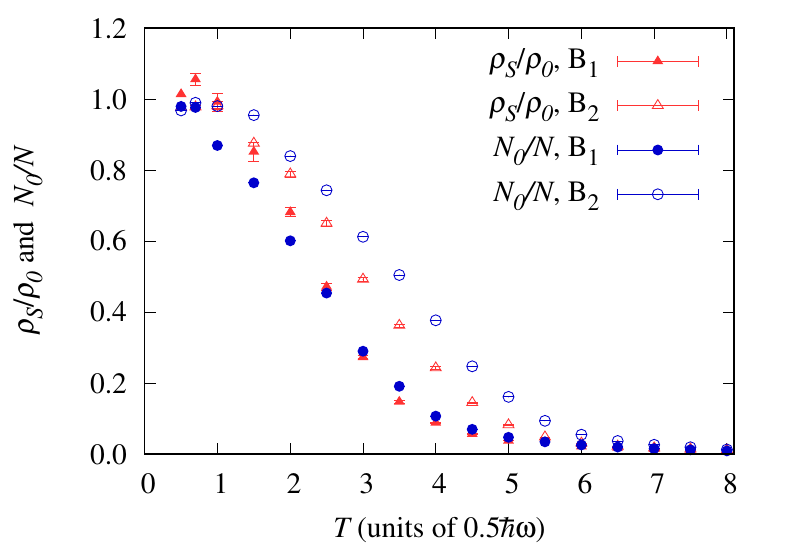}
		\caption{Superfluid and condensate fraction as a function of temperature for $B_1$ and $B_2$ in the shell phase (parameters given in Table \ref{table:scattering_params}). }
		\label{shellsf}
\end{figure}

In Fig \ref{shellsf}, we report the superfluid and condensate fractions as a function of temperature for both components $B_1$ and $B_2$. All reach one in the limit of zero temperature, but the values for the component which is preferably in the outer shell ($B_1$) decrease faster with the temperature, showing a wide regime with a superfluid and condensate fractions smaller than those for the inner species. Again, one observes that at 
temperatures approaching the critical one the condensate fraction becomes larger than the superfluid fraction, particularly for the inner component. 
\section{Summary and conclusions}
Superfluidity and Bose-Einstein condensation of repulsive Bose-Bose mixtures have been explored in the  three phases which they manifest. We observed agreement between DMC and PIMC results at low temperatures, giving us confidence that all possible biases have been eliminated. In both the mixed and the shell regimes, and in the limit of zero temperature, the superfluid fraction, estimated with the area estimator is one, while in the two-blobs regime it depends on the axis with respect to which the rotation is considered. It is below 0.5 when calculated with respect to an axis which is perpendicular to the line passing through two centers of mass. 
Therefore, the two-blobs phase shows an anisotropic superfluid response due to the form in which the system separates. Anisotropy in the superfluid fraction has already been predicted in dipolar gases, where it is caused by the anisotropic character of the interatomic potential~\cite{bombin}, as well as in doped Helium clusters \cite{shinichi2007, sungjin2015}.

 Interestingly, the condensate fraction, which is typically lower than the superfluid fraction, becomes larger than $\rho_s/\rho_0$ when $T$ increases.  The difference between both magnitudes is, in all cases, most pronounced around $T=3.5$  and then it diminishes as both superfluid and condensate fraction drop to zero, around $T=8$.
This effect is also observed in a single species simulation with the same interparticle interaction strength; if it is increased the crossing of condensate and superfluid fraction disappears. 
 We do not expect that our results depend on the model of the interaction potential, as they are obtained in the regime which was previously estimated as universal~\cite{Cikojevic18}.

\section{Acknowledgments}
This work has been supported in part by the Croatian Science Foundation under 
the project number IP-2014-09-2452 and by the by MINECO (Spain) Grant No. 
FIS2014-56257-C2-1-P. 
We acknowledge financial support from Secretaria d'Universitats i Recerca del 
Departament d'Empresa i Coneixement de la Generalitat de Catalunya, co-funded by 
the European Union Regional Development Fund within the ERDF Operational Program 
of Catalunya (project QuantumCat, ref. 001-P-001644).
The computational resources of the Isabella cluster at Zagreb University 
Computing Center (Srce)  were used.

\end{document}